\documentclass[amsmath,amssymb, superscriptaddress, preprint, showpacs]{revtex4-1}
\usepackage[english]{babel}
\usepackage{graphicx}
\usepackage{amsmath}
\usepackage{amssymb}
\usepackage[skins,theorems]{tcolorbox}
\usepackage{amsfonts}
\usepackage{bm}
\usepackage{mathdots}
\usepackage{mathtools}
\usepackage{stmaryrd}
\usepackage{times}
\usepackage{color}
\usepackage{setspace}

\setcitestyle{super}
\setcitestyle{open={},close={)}}
\makeatletter
\renewcommand{\p@subsection}{}
\renewcommand{\p@subsubsection}{}
\renewcommand{\@cite}[1]{\textsuperscript{\,[#1]}}
\makeatother

\begin{document}
%
%
%
%
\newcommand{\va}{{\bf a}} \newcommand{\vA}{{\bf A}} 
\newcommand{\vb}{{\bf b}} \newcommand{\vB}{{\bf B}}
\newcommand{\vc}{{\bf c}} \newcommand{\vC}{{\bf C}}
\newcommand{\vd}{{\bf d}} \newcommand{\vD}{{\bf D}}
\newcommand{\ve}{{\bf e}} \newcommand{\vE}{{\bf E}}
\newcommand{\vf}{{\bf f}} \newcommand{\vF}{{\bf F}}
\newcommand{\vg}{{\bf g}} \newcommand{\vG}{{\bf G}}
\newcommand{\vh}{{\bf h}} \newcommand{\vH}{{\bf H}}
\newcommand{\vi}{{\bf i}} \newcommand{\vI}{{\bf I}}
\newcommand{\vj}{{\bf j}} \newcommand{\vJ}{{\bf J}}
\newcommand{\vk}{{\bf k}} \newcommand{\vK}{{\bf K}}
\newcommand{\vl}{{\bf l}} \newcommand{\vL}{{\bf L}}
\newcommand{\vm}{{\bf m}} \newcommand{\vM}{{\bf M}}
\newcommand{\vn}{{\bf n}} \newcommand{\vN}{{\bf N}}
\newcommand{\vo}{{\bf o}} \newcommand{\vO}{{\bf O}}
\newcommand{\vp}{{\bf p}} \newcommand{\vP}{{\bf P}}
\newcommand{\vq}{{\bf q}} \newcommand{\vQ}{{\bf Q}}
\newcommand{\vs}{{\bf s}} \newcommand{\vS}{{\bf S}}
\newcommand{\vt}{{\bf t}} \newcommand{\vT}{{\bf T}}
\newcommand{\vu}{{\bf u}} \newcommand{\vU}{{\bf U}}
\newcommand{\vv}{{\bf v}} \newcommand{\vV}{{\bf V}}
\newcommand{\vw}{{\bf w}} \newcommand{\vW}{{\bf W}}
\newcommand{\vx}{{\bf x}} \newcommand{\vX}{{\bf X}}
\newcommand{\vy}{{\bf y}} \newcommand{\vY}{{\bf Y}}
\newcommand{\vz}{{\bf z}} \newcommand{\vZ}{{\bf Z}}


\newcommand{\vxi}{{\mbox{\boldmath$\xi$}}}
\newcommand{\vlambda}{\mbox{\boldmath$\lambda$}} \newcommand{\vLambda}{\mbox{\boldmath$\Lambda$}}
\newcommand{\vDelta}{\mbox{\boldmath$\Delta$}}
\newcommand{\vna}{\mbox{\boldmath$\nabla$}}
\newcommand{\vtau}{\mbox{\boldmath$\tau$}}
\newcommand{\vsigma}{\mbox{\boldmath$\sigma$}}
\newcommand{\vepsilon}{\mbox{\boldmath$\varepsilon$}}
\newcommand{\vPsi}{\mbox{\boldmath$\Psi$}}
\newcommand{\vvarphi}{\mbox{\boldmath$\varphi$}}
\newcommand{\vchi}{\mbox{\boldmath$\chi$}}
\newcommand{\valpha}{\mbox{\boldmath$\alpha$}}
\newcommand{\vbeta}{\mbox{\boldmath$\beta$}}
\newcommand{\vgamma}{\mbox{\boldmath$\gamma$}}
\newcommand{\vnu}{\mbox{\boldmath$\nu$}}
\def\vxi{{\mbox{\boldmath$\xi$}}} \def\vlambda{{\mbox{\boldmath$\lambda$}}}
\def\vDelta{{\mbox{\boldmath$\Delta$}}} \def\vna{{\mbox{\boldmath$\nabla$}}}
\def\vtau{{\mbox{\boldmath$\tau$}}} \def\vDelta{{\mbox{\boldmath$\Delta$}}}
\def\vsigma{{\mbox{\boldmath$\sigma$}}} \def\vepsilon{{\mbox{\boldmath$\varepsilon$}}}
\def\vPsi{{\mbox{\boldmath$\Psi$}}} \def\vvarphi{{\mbox{\boldmath$\varphi$}}}
\def\vchi{{\mbox{\boldmath$\chi$}}} \def\valpha{{\mbox{\boldmath$\alpha$}}}
\def\vbeta{{\mbox{\boldmath$\beta$}}} \def\vgamma{{\mbox{\boldmath$\gamma$}}}
%
%
\newcommand{\bdm}{\begin{displaymath}} \newcommand{\edm}{\end{displaymath}}
\newcommand{\nn}{\nonumber}  
\newcommand{\bc}{\begin{center}} \newcommand{\ec}{\end{center}}
\newcommand{\be}{\begin{equation}} \newcommand{\ee}{\end{equation}}
\newcommand{\ra}{{\rm a}} \newcommand{\rb}{{\rm b}} 
\newcommand{\e}{{\rm e}} \newcommand{\rc}{{\rm c}} \newcommand{\rh}{{\rm h}}
\newcommand{\kB}{k_{\rm B}} 
\newcommand{\kF}{k_{\rm F}} \newcommand{\EF}{E_{\rm F}}
\newcommand{\NF}{N_{\rm F}} \newcommand{\pF}{p_{\rm F}} 
\newcommand{\Tc}{T_{\rm c}} \newcommand{\vvF}{v_{\rm F}} 
\newcommand{\vPi}{{\bf\Pi}} \newcommand{\muB}{\mu_{\rm B}}
%
%
\def\b{\textbf} %
\def\<{\left\langle}
\def\>{\right\rangle}
\def\d{\delta}
\def\l{\lambda}
\newcommand{\+}{\dagger}
\def\vnu{{\mbox{\boldmath$\nu$}}}
\def\vxi{{\mbox{\boldmath$\xi$}}}
\def\up{\uparrow}
\def\dn{\downarrow}
\def\Lp{L^{(+)}}
\def\Lm{L^{(-)}}
\def\Mp{M^{(+)}}
\def\Mm{M^{(-)}}
\def\LpC{L^{(+)*}}
\def\LmC{L^{(-)*}}
\def\MpC{M^{(+)*}}
\def\MmC{M^{(-)*}}
\definecolor{DarkRed}{rgb}{0.67,0.335,0.335}
\definecolor{DarkBlue}{rgb}{0.335,0.335,0.67}
\definecolor{LKRed}{rgb}{0.777,0.156,0.164}
\definecolor{LKGreen}{rgb}{0.097,0.535,0.254}
\definecolor{MPIGreen}{rgb}{0.0,0.502,0.502}
\def\tcDR{\textcolor{DarkRed}}
\def\tcDB{\textcolor{DarkBlue}}
\def\tcb{\textcolor{blue}}
\def\tcr{\textcolor{red}}
\def\tcLKR{\textcolor{LKRed}}
\def\tcLKG{\textcolor{LKGreen}}
\def\tcMPIG{\textcolor{MPIGreen}}
%
%
\title{Possible Light-Induced Superconductivity in a Strongly Correlated Electron System}
\author{Nikolaj Bittner}
\altaffiliation{Corresponding author. Current address: Department of Physics, University of Fribourg, 1700 Fribourg, Switzerland}
\affiliation{Max--Planck--Institut f\"ur Festk\"orperforschung, D--70569 Stuttgart, Germany}
\author{Takami Tohyama}
\affiliation{Department of Applied Physics, Tokyo University of Science, Tokyo 125-8585, Japan}
\author{Stefan Kaiser}
\affiliation{Max--Planck--Institut f\"ur Festk\"orperforschung, D--70569 Stuttgart, Germany}
\affiliation{$4^\mathrm{th}$ Physics Institute, University of Stuttgart, D--70569 Stuttgart, Germany}
\author{Dirk Manske}
\affiliation{Max--Planck--Institut f\"ur Festk\"orperforschung, D--70569 Stuttgart, Germany}
\begin{abstract}
{Using a nonequilibrium implementation of the Lanczos-based exact diagonalisation technique we study the possibility of the
light-induced superconducting phase coherence 
in a solid state system after an ultrafast optical excitation. 
In particular, we investigate the buildup of superconducting correlations 
by calculating an exact time-dependent wave function 
reflecting the properties of the system in non-equilibrium 
and the corresponding transient response functions. 
Within our picture we identify a possible 
transient Meissner effect 
after dynamical quenching of the non-superconducting wavefunction 
and extract a characteristic superfluid density that we compare to experimental data. 
Finally, we find that the stability of the induced superconducting state depends crucially 
on the nature of the excitation quench:  
namely, a pure interaction quench induces a long-lived superconducting state, whereas a phase quench leads to a short-lived transient superconductor. 
}
\end{abstract}


\maketitle
\section{I\lowercase{ntroduction}}
Recent developments of ultrashort laser pulses 
allow for optical control of the complex matter on picosecond time-scales. 
Intriguing experiments at mid-IR and THz frequency ranges 
reported controlling non-equilibrium superconductivity:~\cite{Faustietal:2011, 
Huntetal:2015, Kaiseretal:2014, Hu:2014, Huntetal:2016, Mitranoetal:2016, Kaiser:2017}  
Tailored excitation pulses tuned resonantly to specific phonon modes 
have been shown to induce transient superconducting states  
far above equilibrium transition temperature ($T_c$), which 
goes beyond the typical 
photo-doping experiments.~\cite{Pashkinetal:2010, DalConteetal:2012}
On the one hand superconductivity could emerge after suppression of a 
competing order. Here, the key experiment was performed 
via phonon pumping on $1/8$ doped LESCO,~\cite{Faustietal:2011, Huntetal:2015} 
where superconductivity is fully suppressed due to the appearance 
of a competing stripe order stabilized by lattice distortion.
Melting competing stripes with a light pulse allows for the reappearance~\cite{Faustietal:2011, Huntetal:2015} 
and enhancement~\cite{Nicolettietal:2014, Casandrucetal:2015} of the superconductivity. 
This picture 
was also supported by theoretical investigations,~\cite{PatelEberlein:2016, Rainesetal:2015} 
where the \textit{existing} superconductivity was enhanced after suppressing 
competing orders by optical pulses. 
However, most intriguing are experiments that really {\it induce} superconducting coherence 
into the electron system. That allows stabilizing superconducting state 
far above $T_c$;~\cite{Kaiseretal:2014, Hu:2014, Huntetal:2016, Mitranoetal:2016}  
Most prominent in YBCO that was possible throughout the pseudogap phase. 
Key to these experiments is the resonant pumping of 
the $c$-axis infrared-active CuO mode, which basically modulates apical oxygen displacement 
from the CuO planes, leading 
to a redistribution of the interlayer coherence between the planes;~\cite{Hu:2014}  
Although the evidence for the competing order of the incommensurate CDW in YBCO 
was found,~\cite{Foerstetal:2014} the temperature and time dependent 
measurements rules out its melting as a key element of inducing superconductivity.
A leading 
idea emphasizing the role of phonon pumping 
is based on the displacement of the  
lattice structure through non-linear phononics.~\cite{Forstetal:2011, Mankowsky:2014} 
This transient structure frozen at maximum lattice displacement suggests 
an enhanced superconducting coupling strength.~\cite{Subedietal:2014} 
However, comparing with theoretical expectations the transient displacement and 
therefore impact of this effect 
seems rather small to explain the observed enhanced $T_c$. 
Explanations going beyond the structure dynamics 
are focusing on the amplification of the \textit{existing} electron pairing 
either by changing of microscopic parameters responsible 
for superconductivity,~\cite{Sentefetal:2016} by parametric cooling~\cite{Dennyetal:2015} or 
by phononic squeezing.~\cite{Knapetal:2015, Babadietal:2017, Murakamietal:2017, Kennes:2017, Sentef:2017} 
Though, none of them describe the possibility to induce superconducting coherence. 
An important experimental aspect, which was so far disregarded, is the possibility 
of the modulated effective interactions after an optical excitation.~\cite{pomarico2017} 
The significance of such modulations was experimentally shown on 
1D organic molecules,~\cite{Kaiseretal:2014:2, Singlaetal:2015}  
where a controlled change in Coulomb interaction $U$ was achieved, and most spectacular is discussed 
as a key element for possible light-induced superconductivity 
in K-doped C$_{60}$.~\cite{Mitranoetal:2016, Kimetal:2016, MazzaGeorges:2017} 
In the latter case, a transient superconducting 
state was observed after a tailored optical pulse, which modulates the local charge density. 
However, the understanding of the underlying microscopic mechanism 
in nonequilibrium is still missing. 

Following that view we propose here a general concept 
to induce nonequilibrium electronic phase coherence 
after an optical excitation (see Fig.~\ref{fig:Fig1}) 
by emphasizing the role of modulated correlations $U/t_h$. 
In particular, we show the possibility of 
light-induced superconducting coherence following two basic 
concepts: 
pulse stimulated changes in the interaction strength of the electron system 
(e.g. on-site $U$) and modification of the electron hopping $t_h\exp(iA(t))$ 
by the time-dependent vector potential $A(t)$ of the light field. 
To capture the essential physics we simulate 
an electron 
system in nonequilibrium by the extended Hubbard model (EHM) with an additional 
next neighbor interaction $V$ in one dimension (1D) 
at half-filling~\cite{Solyom:1979, Voit:1992, Voit:1995} described by  
the time-dependent Hamiltonian: 
\begin{figure*}[!t]
\centering
\includegraphics[width=0.99\textwidth,clip]{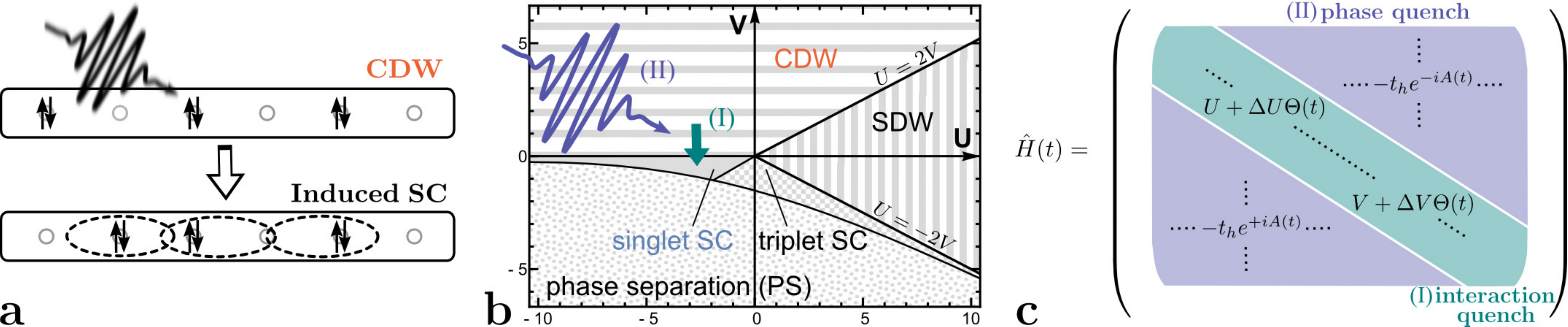} 
\caption{(color online) (a) Inducing phase coherence with an optical pulse. 
Here, spatial distribution of electrons on a lattice before and after pulse corresponds to 
CDW and singlet SC, respectively. 
(b) Illustration of the equilibrium phase diagram for the 1D EHM 
at half filling (T=0) with $U$ and $V$ being the strength of the on-site and nearest neighbor 
interaction from Eq.~\eqref{eq:Ham}, respectively. Each shaded region 
corresponds to a particular phase from EHM. The nonequilibrium transition between 
phases is initiated by an optical pulse following two scenarios: 
(I) an interaction quench or (II) a phase quench. 
(c) Illustration of the time-dependent Hamilton-Matrix. While after interaction quench 
the diagonal part of the Hamiltonian (green) is changed, phase quench modifies its 
off-diagonal part (blue).}
\label{fig:Fig1}
\end{figure*}
 \begin{equation}
 \begin{split}
 \label{eq:Ham}
    \hat{H}(t) &= 
    -t_h \sum_{i \sigma}\left(e^{iA(t)}\hat{c}^\+_{i,\sigma}
    \hat{c}_{i+1,\sigma}+\rm{H.c.}\right)\\
   &+
    \left[U + \Delta U \Theta(t)\right]\sum_i (\hat{n}_{i\up}-\tfrac{1}{2})(\hat{n}_{i\dn}-\tfrac{1}{2})
  + \left[V + \Delta V \Theta(t)\right]\sum_{i}(\hat{n}_i -1)(\hat{n}_{i+1}-1)
  ,
 \end{split}
\end{equation}
where $\hat{c}^\+_{i,\sigma}$ ($\hat{c}_{i,\sigma}$) creates (annihilates) an 
electron with spin $\sigma=\up,\dn$ at site $i$ and  $\hat{n}_{i}=\hat{n}_{i\up}+\hat{n}_{i\dn}$ denotes density operator 
with $\hat{n}_{i\sigma}=\hat{c}^\+_{i,\sigma}\hat{c}_{i,\sigma}$. 
Before the excitation ($t<0$)
we prepare 
the system in equilibrium ground 
state (GS) of the CDW (see Fig.~\ref{fig:Fig1}(a)). 
Then, at around $t=0$ the laser pulse arises, which leads to 
either (I) abrupt changing interaction parameters 
on the diagonal of the Hamiltonian~\eqref{eq:Ham} 
(green part in Fig.~\ref{fig:Fig1}(b) and (c)) or (II)  
to modification of 
the hopping by the time-dependent vector potential $A(t)$ of the light field 
on its off-diagonals (blue part in Fig.~\ref{fig:Fig1}(b) and (c)). 
We note that these excitations are tailored to dominantly change the effective correlation in case (I) (as also showed for the same system by DMRG calculations~\cite{Paeckel}) and to effective transfer of the pulse energy to electrons in case (II).~\cite{Luetal2012} Hence, one drives 
the system out of equilibrium 
and can induce transient superconducting state (Fig.~\ref{fig:Fig1}(a)). 

\section{M\lowercase{ethod}}
In case (I) one deals with the quench protocol,~\cite{Ecksteinetal:2010} where at 
$t=0$ the interaction parameters $U$ and $V$ are abruptly changed by $\Delta U$ and $\Delta V$. Recently this type of quenching was experimentally realized in 1D organic molecules.~\cite{Kaiseretal:2014:2, Singlaetal:2015}
The situation (II) represents excitation with a light pulse 
with a finite duration time.  
The pulse is included into the Hamiltonian 
by means of the Peierls substitution and is modelled 
by a time-dependent vector potential 
 $A(t)=A_0 e^{-t^2/2\tau^2}\cos\left(\omega_{\text{pump}}t\right)$,
with the amplitude  $A_0$, the frequency $\omega_\mathrm{pump}$, and 
the full width at half maximum $\tau$. 
The temporal evolution of the system obeys the Schr\"odinger equation, 
which we numerically solve by employing at each time step $t+\d t$ Lanczos method generating a tridiagonal matrix 
with the eigenvectors $\left.|\phi_l\>$ and corresponding eigenvalues $\epsilon_l$. 
These results are subsequently used to approximate the time-dependent wave function:~\cite{Luetal2012}
\begin{equation}
\hspace{-0.02cm}\left.|\psi(t\hspace{-0.1cm}+\hspace{-0.1cm}\d t)\>\hspace{-0.1cm}\approx\hspace{-0.1cm}e^{-i\hat{H}(t)\d t}
 \hspace{-0.1cm}\left.|\psi(t)\>\hspace{-0.1cm}\approx
 \hspace{-0.15cm}\sum_{l=1}^M\hspace{-0.1cm}e^{-i\epsilon_l\d t}|\hspace{-0.1cm}\left.\phi_l\>
 \hspace{-0.1cm}\<\phi_l|\psi(t)\>\ .
\label{eq:WFt}
\end{equation}
Based on the knowledge of $\left.|\psi(t)\>$ we calculate  
correlation functions as well as optical conductivity, which are good quantities to examine 
a contrast between two neighboring CDW and SC phases. 
The 
temporal evolution of the charge order can be traced by the time-dependent 
density-density correlation function:
\begin{equation}
\label{eq:C}
 	C(j,t)=\frac{1}{L}\sum_{l=0}^{L-1}\<\psi(t)|\hat{n}_{l+j}\hat{n}_l|\psi(t)\>\ ,
\end{equation}
at lattice site $j$ with lattice size $L$. To investigate the dynamics 
of the singlet superconducting correlations we introduce the on-site correlation function:
\begin{equation}
\label{eq:P1}
  P_1(j,t)=\frac{1}{L}\sum_{l=0}^{L-1}\<\psi(t)\left|\hat{c}_{l+j\dn}^\+\hat{c}_{l+j\up}^\+
  \hat{c}_{l\up}\hat{c}_{l\dn}\right|\psi(t)\>\ . 
\end{equation}
The optical pump-probe conductivity $\sigma(\Delta t, \omega)$, 
we calculate by using an additional weak probe pulse described by 
$A_\mathrm{pr}(\Delta t, t)$ at time delay $\Delta t$.~\cite{shao2016,zala2014} 
For a given pump we then sequentially calculate the time-dependent current density without and with the 
probe pulse. 
The difference in both results gives the current density 
$j_\mathrm{pr}(\Delta t, t)$ induced by the probing excitation. Finally, the 
optical conductivity is calculated from the Fourier 
transformation of $j_\mathrm{pr}(\Delta t, t)$ and 
the $A_\mathrm{pr}(\Delta t, t)$ with respect to $t$:
\begin{equation}
\label{eq:OptCond}
 \sigma(\Delta t, \omega)=\frac{j_\mathrm{pr}(\Delta t, \omega)}
 {i\left(\omega + i\eta \right)L A_\mathrm{pr}(\Delta t, \omega)}\ .
\end{equation}
For the broadening of the spectral lines we added artificially a small 
number $\eta=1/L$. 

For the numerical calculations, we use unless otherwise specified 10-site lattice with periodic boundary conditions and 
set hopping parameter $t_h=1$. Further, we measure the energy and time 
in units $t_h$ and $t_h^{-1}$, respectively. For the Lanczos method we take $M=40$ iterations and the time interval $\d t=0.01$.

\section{R\lowercase{esults}}

\subsection{Induced phase coherence due to interaction quench}
\begin{figure}[!b]
 \centering
  \includegraphics[width=1.0\columnwidth,clip]{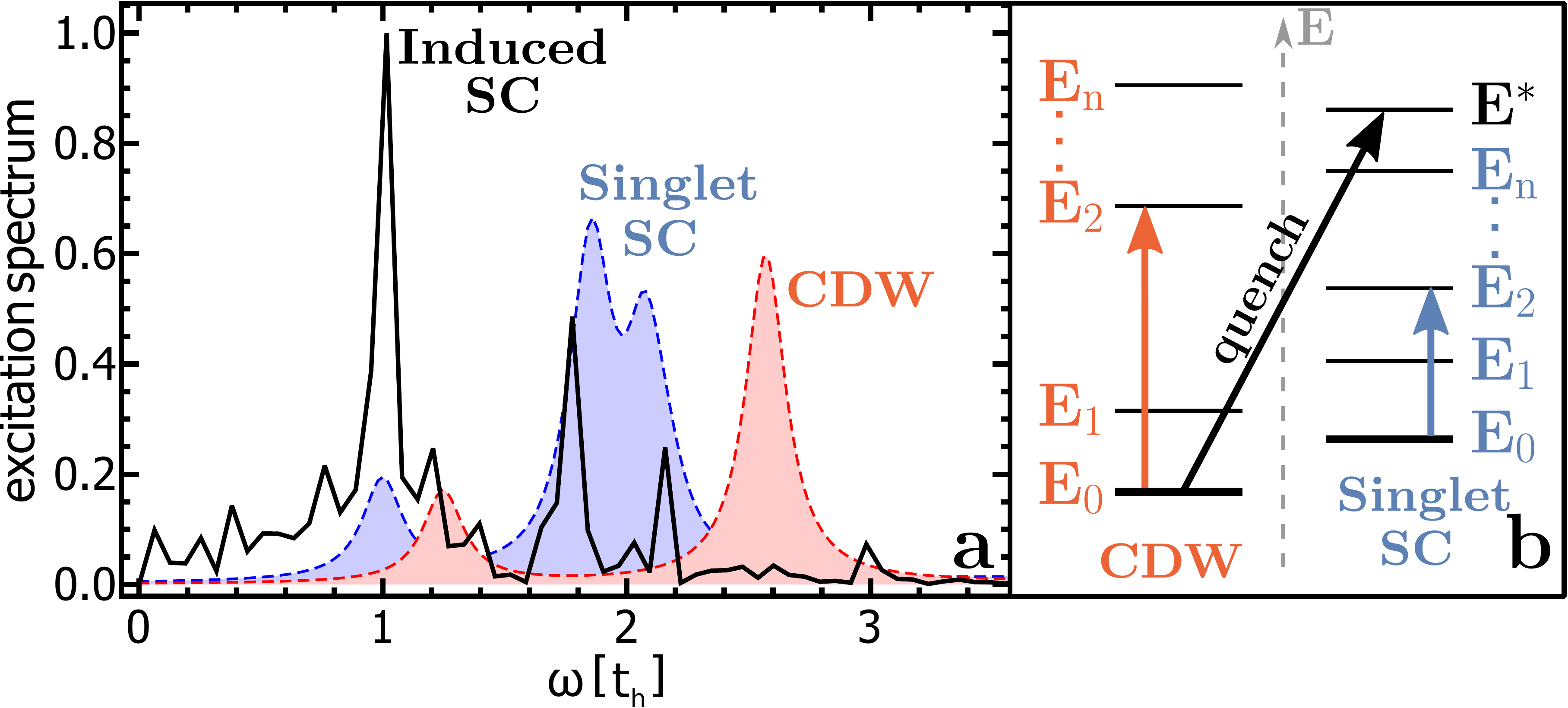}
\caption{(color online) (a) Excitation spectrum of the system in nonequilibrium 
(black solid line) compared with the spectra in equilibrium 
SC phase (blue region) and equilibrium CDW phase (red region). 
(b) Schematic illustration of the interaction quench together with excitation processes 
in SC and CDW phases. Black bars represent energy levels $E_n$ in both phases. 
A nonequilibrium state is indicated by $E^*$. 
}
\label{fig:Fig2}
\end{figure}

Light-induced abrupt change in the interaction strength of the electron system 
leads to the modification in its eigenstates and, 
in turn, to the redistribution of the electrons on a lattice. 
For visualization we have calculated 
the excitation spectrum of the system in nonequilibrium 
and compared it with its equilibrium counterparts before and after quenching 
(see Fig.~\ref{fig:Fig2}(a)). 
At $t<0$ we prepare the system in equilibrium GS of the CDW with $U=-4, V=0.25$ 
and switch abruptly at $t=0$ the 
interaction parameter $V$ into the superconducting 
region with $V=-0.25$ (green part in Fig.~\ref{fig:Fig1}).
Analyzing the excitation spectrum in nonequilibrium (black line) in Fig.~\ref{fig:Fig2}(a) 
we can draw some important conclusions. 
First of all, there are no peaks in spectrum, which could be identified with 
the equilibrium CDW phase (red area). 
Moreover, the most intensive peaks in spectrum correspond 
to the low-energy excited states in the equilibrium SC phase (blue area). 
Finally, since the interaction quench is a nonadiabatic 
process, the system does not appear in GS 
of the singlet SC phase after quenching, but in some nonequilibrium state $E^*$. 
Physically that means, that after quenching the system 
goes directly into a nonequilibrum state in the SC phase and the CDW is fully suppressed, 
as schematically summarized in Fig.~\ref{fig:Fig2}(b).  
Moreover, this tailored excitation dominantly leads to an effective transfer of the applied energy to the changes of the correlation in the system, rather then 
 to its heating.~\cite{Paeckel} As shown below, the SC correlations in nonequilibrium are almost immediately induced after 
quenching (within few $t_h^{-1}$). Otherwise, 
the system would end up 
in a thermal state with 
a high effective temperature, where the superconducting correlations are not expected. Note that also experimentally no change of the kinetic energy was observed during interaction quench in 1D organic molecules.~\cite{Kaiseretal:2014:2}

\begin{figure}[!b]
\centering
%
\includegraphics[width=1.0\columnwidth,clip]{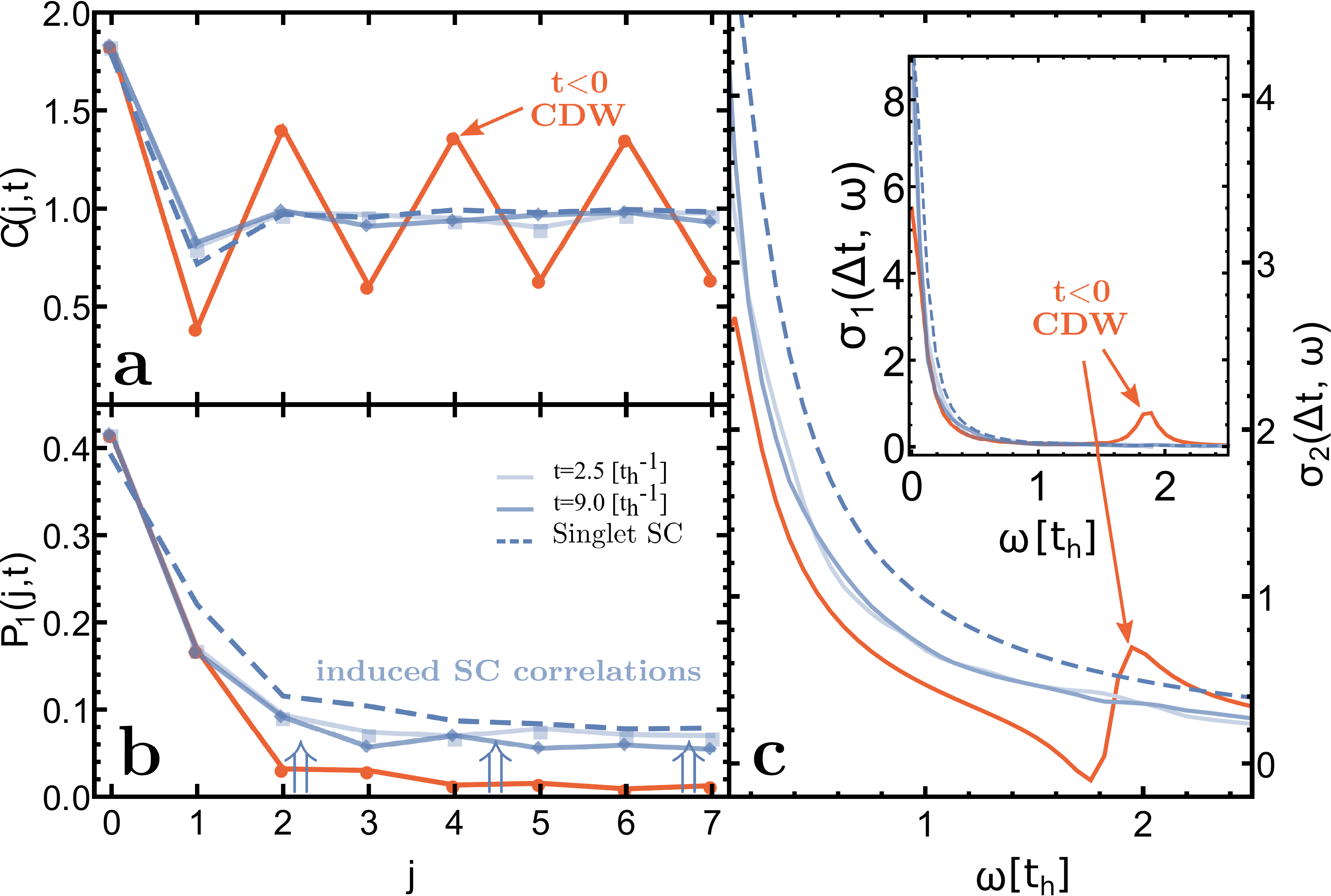}
\caption{(color online) Time-dependent (a) density-density and (b) 
on-site correlation 
functions for 14-site lattice obtained for the interaction quench 
($U=-4$, $V=0.25$, $\Delta V=-0.5$)
at times: $t<0$ (red solid line), 
$t=2.5$ (light blue solid line) and $t=9$ (blue solid line). 
Results for 
equilibrium SC phase are indicated by blue dashed lines. 
(c) The calculated corresponding imaginary optical conductivity $\sigma_2(\Delta t, \omega)$. 
Its real part $\sigma_1(\Delta t, \omega)$ is shown in inset.
}
\label{fig:Fig3}
\end{figure}
To show how superconducting signatures emerges after interaction quench with time, we calculate 
correlation functions and more significantly the experimentally 
relevant time-dependent optical conductivity $\sigma(\Delta t, \omega)$ 
(see Eq.~\eqref{eq:OptCond}). 
First, we plot in Fig.~\ref{fig:Fig3}(a) and (b) time-dependent density-density $C(j,t)$ and 
on-site $P_1(j,t)$ correlation functions defined in Eqs.~\eqref{eq:C} and~\eqref{eq:P1}, respectively.
Clearly, the CDW state prepared at $t<0$ and 
showing characteristic "zigzag" structure in Fig.~\ref{fig:Fig3}(a) 
is strongly suppressed after quenching. 
At the same time, $P_1(j,t)$ presented in Fig.~\ref{fig:Fig3}(b) 
shows strong enhancement of the superconducting correlations in nonequilibrium. 
Moreover, a direct comparison of $P_1(j,t)$ in nonequilibrium 
with the result for equilibrium 
superconducting GS (blue dashed line) reveals a quite good agreement. 
Finally, we focus on the 
optical conductivity. 
Superconductivity is identified in the complex $\sigma(\Delta t, \omega)$ as follows: 
Due to the perfect diamagnetic response of a superconductor 
the imaginary part $\sigma_2(\Delta t, \omega)$ 
diverges with a characteristic Landau $1/\omega$-behavior for $\omega\to0$ 
indicating the Meissner effect. 
The perfect conductivity leads in the real part $\sigma_1(\Delta t, \omega)$ 
to the opening of a superconducting gap of the optical conductivity and 
a transfer of the spectral weight into the zero frequency $\delta$-peak. 
In Fig.~\ref{fig:Fig3}(c) we show $\sigma_2(\Delta t, \omega)$ and $\sigma_1(\Delta t, \omega)$ (inset) 
calculated in equilibrium superconducting state (blue dashed line) and 
after quenching (solid lines). 
Clearly, after quenching $\sigma_2(\Delta t, \omega)$ shows disappearance of the CDW response visible as clear spectral feature 
$\omega\approx 1.8$ in the equilibrium spectra (red solid line) and the appearance of a clear inductive response 
with a superconducting-like $1/\omega$ behavior in the transient spectra (blue solid lines). 
That indicates the emergence of the superconducting correlations via the Meissner effect. 
In order to characterize the induced superfluid density we calculated $\<\omega\Delta\sigma_2\>$.~\cite{SM0} Here, we found a sudden enhancement of its value after the quenching with the subsequent oscillations. Moreover, the transient value of $\<\omega\Delta\sigma_2\>$ is comparable with its equilibrium counterpart for the superconducting phase, if the temperature due to the quench is taken into account.
Equivalent indications were also found in the temporal evolution of 
$\sigma_1(\Delta t, \omega)$, 
where the spectral weight of the low energy peak at $\omega\approx 1.8$ 
corresponding to the equilibrium CDW state is shifted after 
quenching into the low-energy peak at $\omega\approx0$.~\cite{SM4}  
We note that the finite low frequency tail in the equilibrium CDW state 
is an artefact of the finite-size calculation. 
In the induced non-equilibrium case all spectral weight that gets transferred into the low frequency peak is interpreted as spectral weight that forms the transient $\delta$-peak of the superfluid. Note that due to the finite size of our model system all this weight is below the low frequency cut off frequency. This weight tracks the equilibrium superconducting $\delta$-peak shown as blue dashed line in the inset of Figure 3c that also appear with finite width in our finite size model.
\subsection{Realistic phase quench and comparison with experiment}
\begin{figure}[!b]
\centering
\includegraphics[width=1.0\columnwidth,clip]{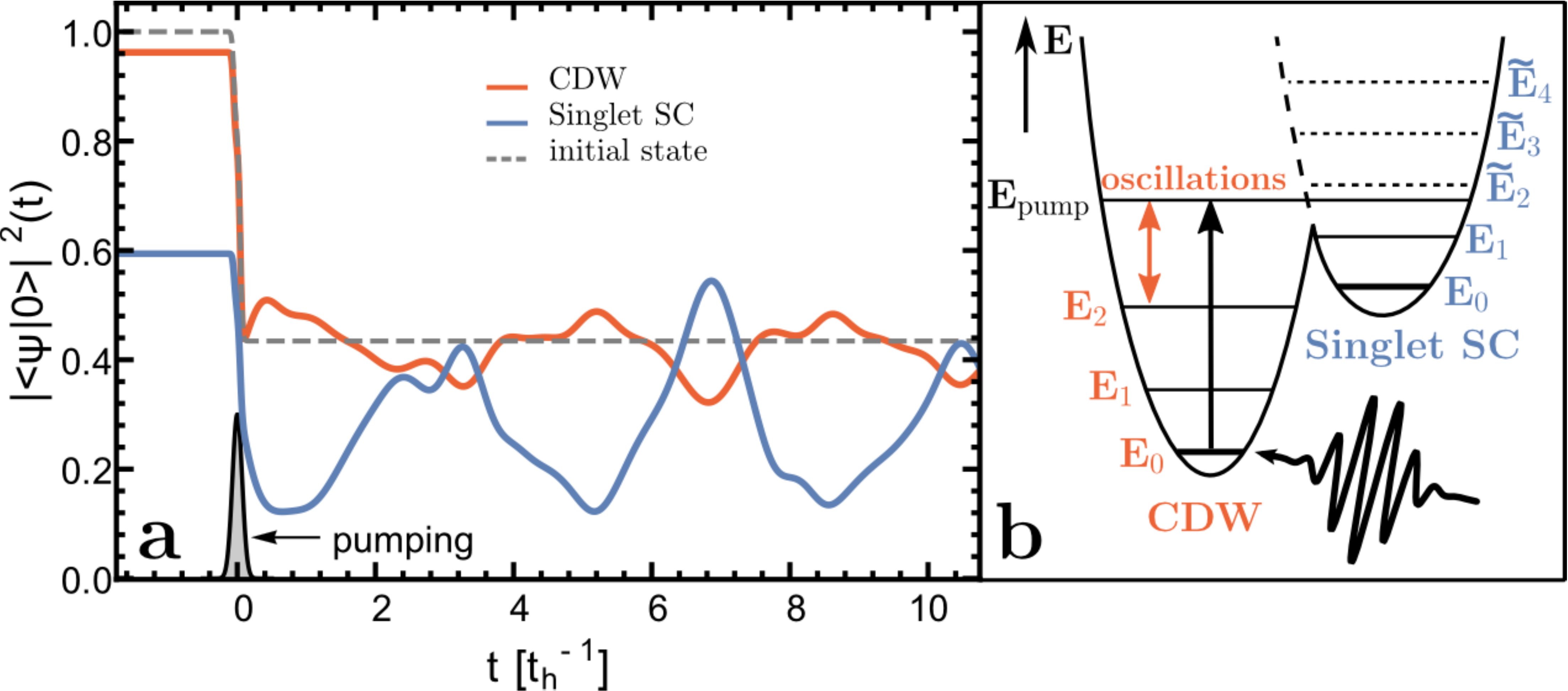}
\caption{(color online) 
(a) Time-dependent overlap function $|\<\psi(t)|0\>|^2$ for: 
$U=-3$ and $V=0.7$ (CDW), $-0.5$ (Singlet SC), and $0.5$ (initial state). 
The pulse has a Gaussian shape with 
parameters:  $A_0$=5, $\omega$=2.38, $\tau$=0.05
and is indicated by the black region around $t=0$. 
(b) Illustration of the excitation of 
a system by
a strong light pulse.}
\label{fig:Fig4}
\end{figure}

Now, we investigate the second important case, where a light pulse 
modifies the electron hopping with the possibility to induce phase coherence 
between the electron pairs on the lattice. 
Since it is technically impossible to calculate the superconducting order parameter within our method, 
we use for our studies of the induced superconducting correlations some related quantities.
First, we calculate 
the projection of the 
time-dependent wave function $\left.|\psi(t)\>$ from Eq.~\eqref{eq:WFt} to different ground states 
$\left.|0\>$ (see Fig.~\ref{fig:Fig4} (a)). 
The system is initially prepared 
in the GS with $U=-3, V=0.5$ (CDW phase) 
and at $t=0$ 
gets exposed to the light field with a finite width 
(blue part in Fig.~\ref{fig:Fig1}). 
The central frequency of the pulse is resonantly tuned to the first low-energy excited state 
in the CDW phase to guarantee an effective transfer of the pulse energy to electrons.~\cite{Luetal2012} 
Its duration time is 
set to 
$\tau=0.05$ allowing  
to reach simultaneously different excited states of the system. 
Clearly, 
after pumping the CDW component in $\left.|\psi(t)\>$
is only partially suppressed (red solid line), whereas 
the projection to the superconducting GS 
(blue solid line) shows a temporal 
enhancement in the overlap function $|\<\psi(t)|0\>|^2$. 
One should also note that the overlap of the initial CDW state with the superconducting one before quench is nonzero and indicates non-orthogonal ground states.
However, 
at $t\approx 7$ 
the overlap with the superconducting GS  
is even larger, 
than with the initial state (gray dashed line). 
This might indicate excitation of the system 
to some joint state of the CDW and SC, 
as illustrated in Fig.~\ref{fig:Fig4}(b). 
The interpretation of a dynamical coexistence of both phases 
is also 
supported by the excitation spectrum and correlation functions.~\cite{SM1} 
Note that owing to the resonant driving of the system the pulse energy goes effectively to the changes of correlations.
Otherwise, the corresponding thermal state would have rather high effective temperature, where  
the buildup of the superconducting correlations is not expected.

\begin{figure*}[!t]
\centering
%
\includegraphics[width=0.8\textwidth,clip]{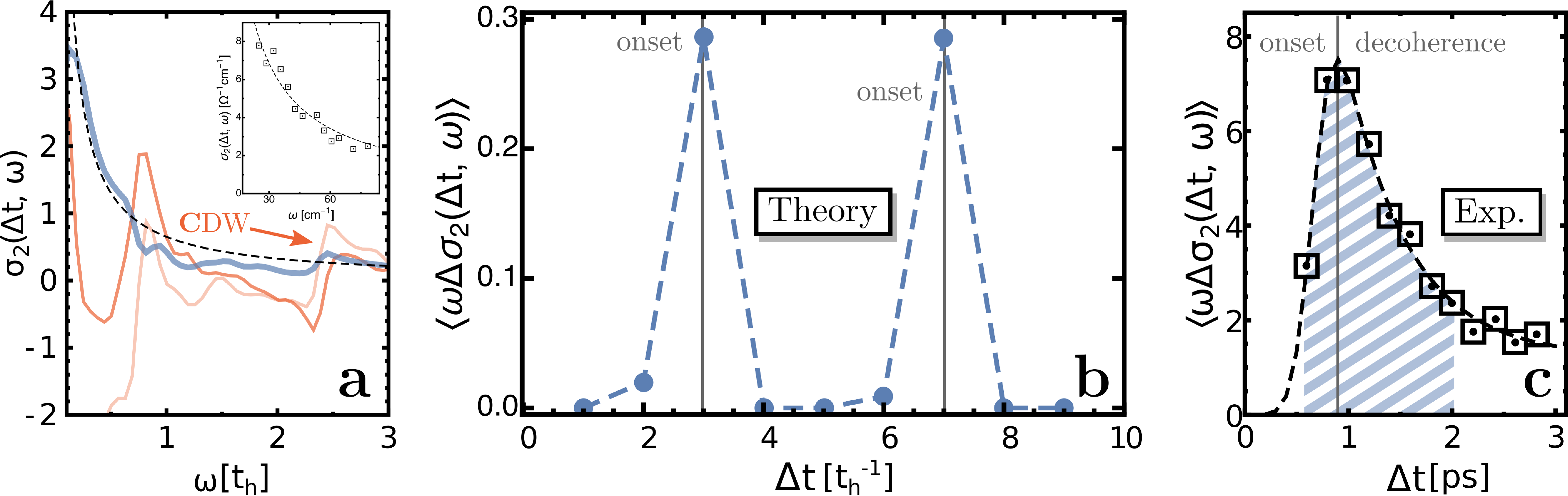}
\caption{(color online) (a) Time-dependent imaginary conductivity $\sigma_2(\Delta t, \omega)$. 
Theoretically obtained results at different time delays are presented by the solid lines: 
$\Delta t$=1 (light red), $\Delta t$=2 (red), and $\Delta t$=3 (blue). 
The $1/\omega$-fit for the latter case is shown by the dashed line. 
Experimentally measured data for YBCO 6.45 at $T=100$K and $\Delta t=0.8$ps is illustrated by the black squares in inset. 
(b) Theoretically calculated for the Hubbard model and 
(c) experimentally measured~\cite{Huntetal:2016} for YBCO 6.45 at $T=100$K 
transient value of $\<\omega\Delta\sigma_2(\Delta t, \omega)\>$  
for different time delays $\Delta t$. The gray solid lines separate 
onset and decoherence areas on figures. 
}
\label{fig:Fig5}
\end{figure*}
Next, to explore the temporal dynamics of this joint state and 
to find further fingerprints 
of the superconducting phase coherence, we calculate the time-dependent 
optical conductivity (see Fig.~\ref{fig:Fig5}). In Fig.~\ref{fig:Fig5}(a) 
we show first $\sigma_2(\Delta t, \omega)$,~\cite{SM2}  where 
we observe a reduction of the equilibrium 
low-energy CDW response at both $t=1$ (light red line) and $t=2$ (red line)
with the appearance of an in-gap state at $\omega\approx 0.7$. 
Subsequently, at $t=3$ (blue line) the response from CDW is disappeared and the data can be fitted with $1/\omega$ (black dashed line).
This characteristic Landau $1/\omega$-like behavior we attribute 
as the diamagnetic response emerging from the superconducting correlation 
due to the Meissner effect. 
We note that a $1/\omega$ behavior is a generic characteristics  
found throughout induced superconductors.~\cite{Faustietal:2011, Nicolettietal:2014, Kaiseretal:2014,Mitranoetal:2016} 
For illustration, we plot in inset to Fig.~\ref{fig:Fig5} the induced inductive response  
exemplary for the cuprate YBCO 6.45 under phonon pumping~\cite{Huntetal:2016} 
(open black squares). 
These experimental data shows an induced $1/\omega$ behaviour and was also interpreted as an induced inductive response of a transient Meissner effect. 
Finally, to characterize the induced superfluid density we calculate 
$\omega\Delta\sigma_2(\Delta t, \omega)$ by fitting the 
$1/\omega$ divergence at low frequencies for each spectrum. 
Fig.~\ref{fig:Fig5}(b) shows the extracted transient superfluid density as a function 
of time delay $\Delta t$. Clearly, pumping the system 
leads to an induction of a finite   
superfluid density with an onset between $\Delta t=1$ to 3. 
On the other hand, the dynamical coexistence of the CDW 
and superconducting excitations leads to the characteristic oscillating behavior discussed 
above. This is also seen in the reappearing onset in $\omega\Delta\sigma_2(\Delta t, \omega)$ 
between $\Delta t=5$ to 7. 
Hence, the transient character of $\left.|\psi(t)\>$ 
(see Fig.~\ref{fig:Fig4}(a)) describes a periodic decoherence and therefore leads to   
a short-lived transient superconducting state. 
While the onset behaviour of the above mentioned cuprate YBCO 6.45 (see Fig.~\ref{fig:Fig5}(c)) shows clear similarities to our model calculations the stability of the light induced state has to be interpreted differently: The short-lived character of the experimentally induced state is due to dephasing of the light-induced superfluid by the loss of long range coherence into several decay and deposing channels,~\cite{Huntetal:2016} while in our model calculation the short lived character is due to the oscillatory behaviour between SC and CDW order in the model system without decay channels.
\section{C\lowercase{onclusion}}
In the present paper we have studied the possibility that 
tailored light pulses that modulate correlations induce  
superconducting phase coherence in transient nonequilibrium states. 
The nonequilibrium dynamics in a solid state system we simulated using time-dependent Lanczos-based exact diagonalization technique.

Calculating the transient ac-optical conductivity we found transferring of the spectral weight into the zero frequency peak in $\sigma_1$ indicating increase of the conductivity of the system. More importantly we observed within the low-frequency cut-off given by the dimensionality of our system $1/\omega$ divergence at $\omega\to0$ in  imaginary conductivity $\sigma_2$. In contrast to a conventional conductor, where one finds a Drude peak in $\sigma_1$ and a broad peak in $\sigma_2$ defined by the scattering rate, these signatures indicate a superconducting state. However, the optical properties of a superconductor cannot be distinguished from that of a perfect conductor. This is of particular importance, since the electron system that we consider here is dissipationless. In this case the 1/omega behavior in $\sigma_2$ would not be a unique identification for the Meissner-effect. Therefore, to exclude the possibility of the photo-doping of the system, where the photo-doped electrons would lead to a metallic response, we calculated the double occupancy function:~\cite{SM3}  here we find the existence of paired electrons in the driven state, which is indicative for a superconductor. Metallic response would show instead single occupancies. We note that a similar dependence of the double occupancy function was observed for the
Falicov-Kimball model and was attributed to the melting of CDW order.~\cite{Freericksetal:2017} However, the response of the CDW in our case in a   nonequilibrium state is almost completely suppressed. 
Moreover, both in the correlation functions, excitation spectra and in the transient wavefunction we find 
further fingerprints
for the transient superconducting response in the light induced state. 
Hence, our comprehensive picture allows interpreting the $1/\omega$
inductive response as a transient Meissner effect, since it clearly emerges from 
the induced superconducting correlations and excitations. The behavior of $\sigma_2$ 
allows us extracting the induced superfluid densities and a comparison with 
experimental quantities. 
Based on the quench processes the induced nonequilibrium states show distinct 
different transient behaviors.  
While for a pure interaction quench 
a long-lived superconducting state can be induced, 
for a phase quench a short-lived transient superconducting state emerges. 

\begin{acknowledgments}
The authors are grateful to A.~Schnyder, A.~Cavalleri, Y.~Murakami, 
and H.~Krull for helpful discussions. N.B. thanks the YITP in Kyoto, Japan for its hospitality. S.K. acknowledges support from the Ministerium f\"ur Wissenschaft, Forschung und Kunst Baden-W\"urttemberg through the Juniorprofessuren-Programm and support from the Daimler und Benz Stiftung. 
\end{acknowledgments}
\bibliographystyle{apsrev4-1}
\bibliography{BTKM}
\newpage
\begin{center}
\large{\bf  Possible Light-Induced Superconductivity in a Strongly Correlated Electron System\\
{\bf Supplement material}
}\\
\ \newline
\normalsize{Nikolaj Bittner$^1$, Takami Tohyama$^2$, Stefan Kaiser,$^{1,3}$ and Dirk Manske$^1$\\
$^1${\it Max--Planck--Institut f\"ur Festk\"orperforschung, D--70569 Stuttgart, Germany}\\
$^2${\it Department of Applied Physics, Tokyo University of Science, Tokyo 125-8585, Japan}\\
$^3${\it $4^\mathrm{th}$ Physics Institute, University of Stuttgart, D--70569 Stuttgart, Germany}
}\\
\end{center}
\twocolumngrid
\singlespacing
\section*{A. Interaction quench}

\renewcommand*{\thefigure}{S.\arabic{figure}} 
\setcounter{figure}{0}

\begin{figure}[!htb]
\centering
%
\includegraphics[width=0.85\columnwidth,clip]{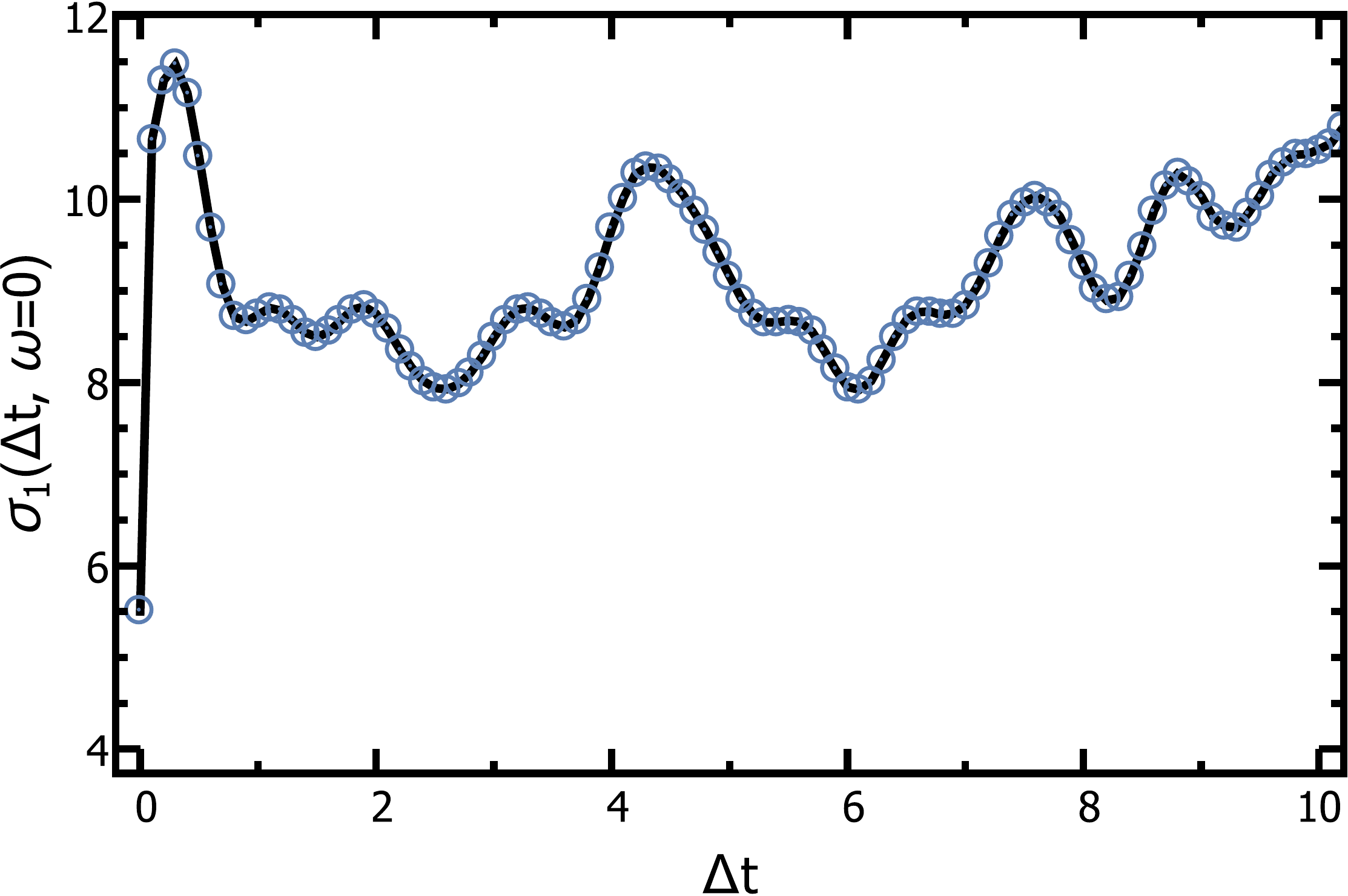}
\caption{(color online) Temporal evolution of the magnitude of the low energy peak ($\omega$ = 0) in the $\sigma_1(\Delta t, \omega)$ spectrum after an interaction quench.
}
\label{fig:LEP}
\end{figure}

\textit{Quench induced enhancement of the low energy peak in $\sigma_1$}. 
In Fig.~\ref{fig:LEP} we plot the temporal evolution of the magnitude of the low energy peak measured at frequency $\omega=0$ in the real part of the optical conductivity spectrum ($\sigma_1(\Delta t, \omega=0)$).
Clearly, after quenching there is a sudden enhancement of $\sigma_1(\Delta t, \omega=0)$ with the subsequent oscillations. At the same time, as discussed in the main text in connection to Fig.3, the CDW peak at $\omega\approx 1.8$ is completely disappeared. Thus, we can conclude that the spectral weight of the CDW peak is shifted into the low energy peak at $\omega=0$.

\begin{figure}[!htb]
\centering
%
\includegraphics[width=0.85\columnwidth,clip]{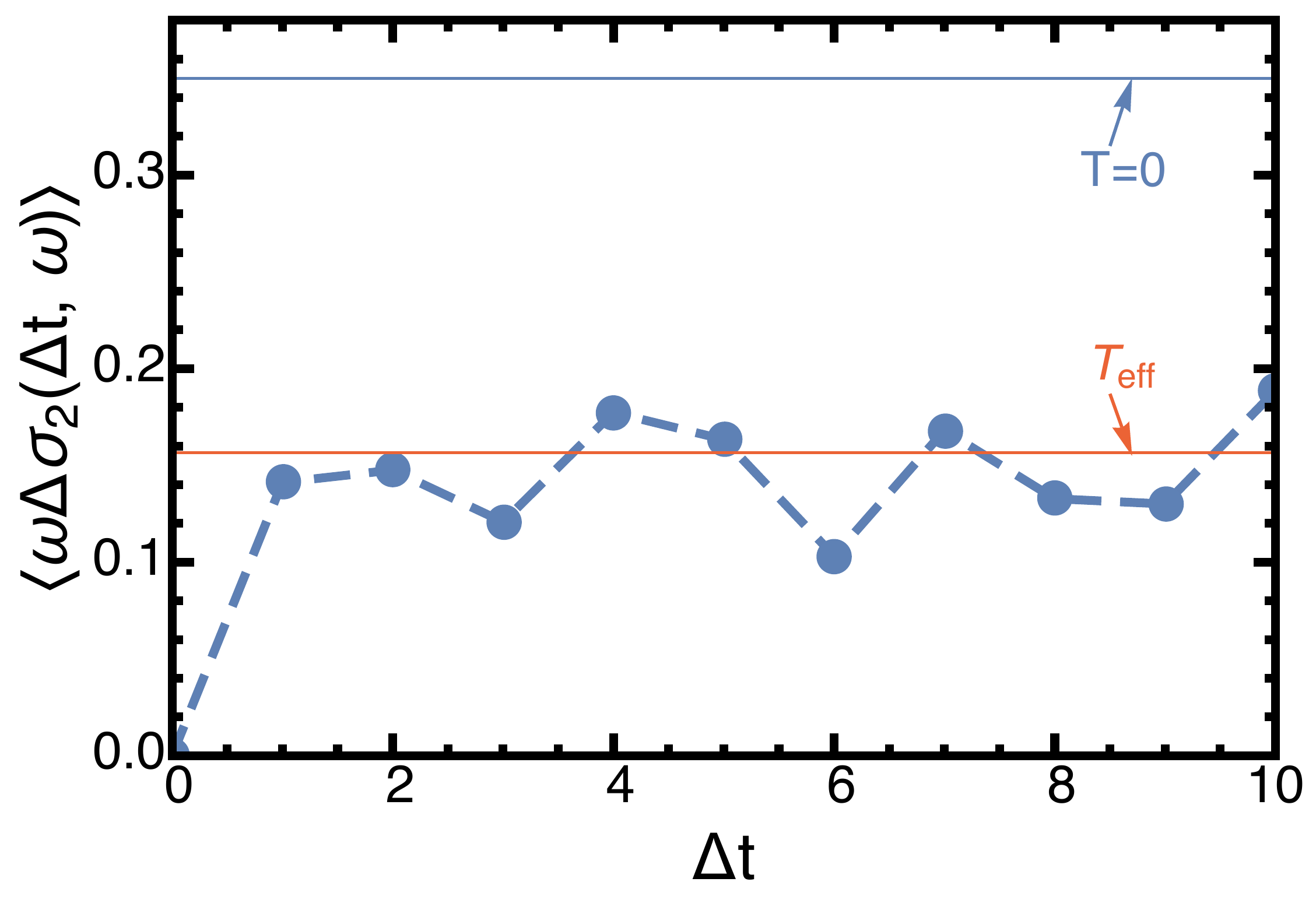}
\caption{(color online) Transient value of $\<\omega\Delta\sigma_2(\Delta t, \omega)\>$ after an interaction quench (blue dashed line). The parameters of the quench are the same, as in the main text. The blue solid line corresponds to its equilibrium counterpart $\<n\>_\mathrm{sc}$ in the superconducting phase ($U=-4, V=-0.25$) at $T=0$. The red solid line represents the value of $\<\omega\Delta\sigma_2\> = \<n\>_\mathrm{sc} \sqrt{1-T_\mathrm{eff}/T_\mathrm{c}}$ with the numerically estimated ratio $T_\mathrm{eff}/T_\mathrm{c}=0.8$.
}
\label{fig:ws2}
\end{figure}
\textit{Quench induced enhancement in $\omega\Delta\sigma_2$}. For characterization of the induced superfluid density we calculated $\<\omega\Delta\sigma_2\>$ by fitting the $1/\omega$ divergence at low frequencies for each spectrum. We note that due to the finite size of our model we corrected $\sigma_2$ from the CDW contribution, i.e. $\Delta\sigma_2(\Delta t,\omega)=\sigma_2(\Delta t,\omega) - \sigma_2^\mathrm{eq,CDW}$. The results are plotted in Fig.~\ref{fig:ws2} by the blue dashed line. As one can see from the figure, $\<\omega\Delta\sigma_2\>$ shows a sudden enhancement after the quenching with the subsequent oscillations. This temporal behavior demonstrates dynamics  reminiscent of $\sigma_1(\Delta t,\omega=0)$ (cf. Fig.~\ref{fig:LEP}). Moreover, we compared the value of $\<\omega\Delta\sigma_2\>$ with its equilibrium counterpart $\<n\>_\mathrm{sc}$ at $T=0$ for the superconducting phase with $U=-4, V=-0.25$ and $\<n\>_\mathrm{sc}\sqrt{1-T_\mathrm{eff}/T_\mathrm{c}}$, which are shown in Fig.~\ref{fig:ws2} by the blue and red solid lines, respectively. 
In the latter case we numerically estimated the ratio $T_\mathrm{eff}/T_\mathrm{c}=0.8$. 
One can see that the temporal spectral weight of $\omega$ times $\sigma_2$ shows almost the same or even slightly larger values (at $\Delta t\approx 4, 7$ and 10) then in  equilibrium, if the temperature due to the quench is taken into account ($T_\mathrm{eff}$). 

\addtocounter{figure}{1}
\begin{figure}[!b]
\centering
%
\includegraphics[width=0.85\columnwidth,clip]{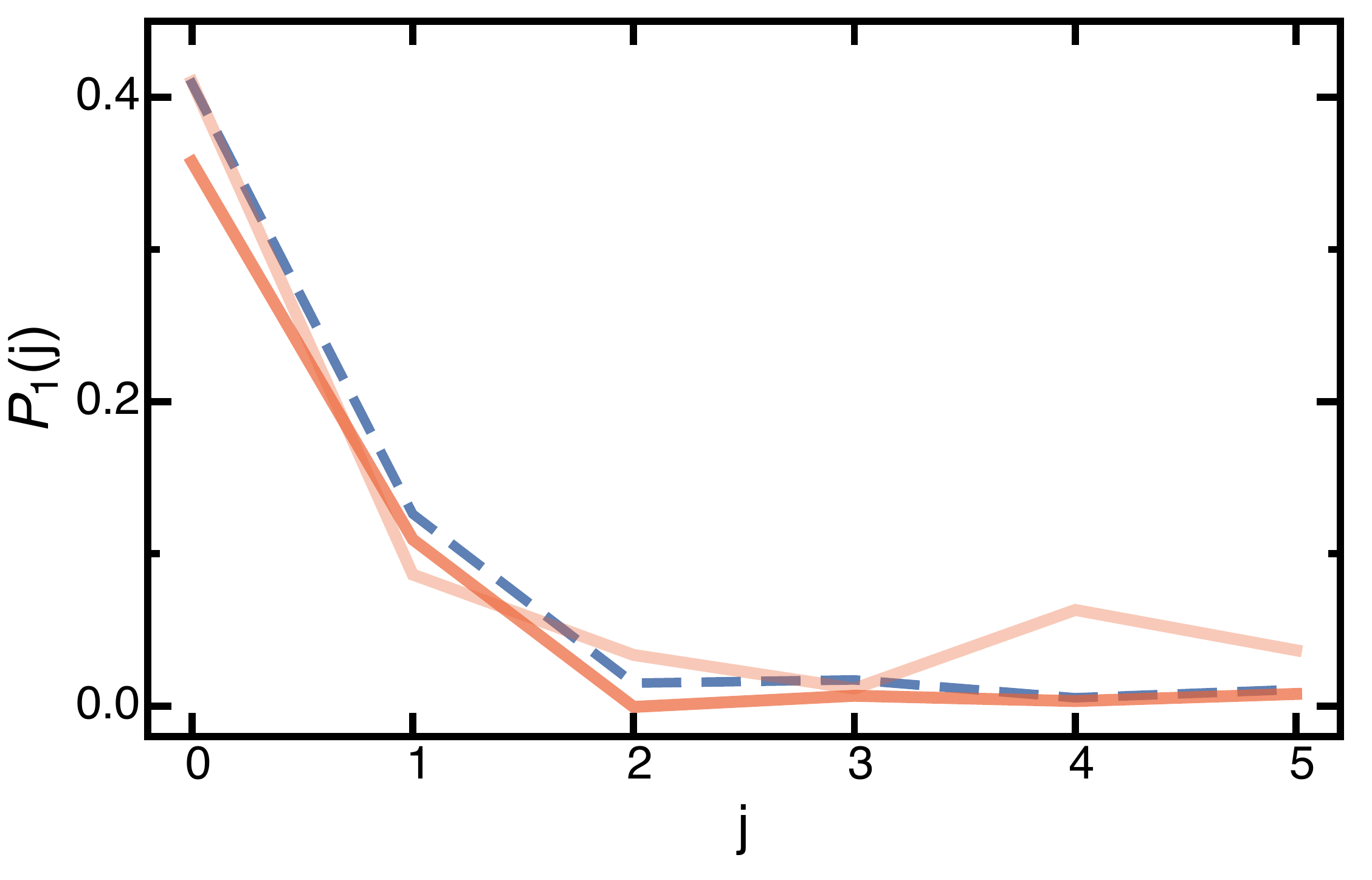}
\caption{(color online) On-site correlation function versus lattice distance $j$ calculated for $U=-3$ and $V=0.5$.  The calculations are done exemplary for the first 3 eigenstates: ground state with $\Delta E$=0 (blue dashed line), $\Delta E_1=1.6295$ (light red solid line), and $\Delta E_2=2.5479$ (red solid line). \quad\quad\ 
}
\label{fig:P1ExcSt}
\end{figure}

\textit{Cluster size dependence.} Here, we compare the results of calculations for 10 and 14-sites lattice after identical quench excitation. We use the same quench protocol as described in the main text: the system is initially prepared in the CDW state with $U=-4$ and $V=0.25$. At $t=0$ the next-neighbour interaction strength is then suddenly changed by $\Delta V=-0.5$. The time-dependent results for density-density $C(j,t)$ and on-site $P_1(j,t)$ correlation functions are shown in Fig.~\ref{fig:scaling}. Clearly, for both 10 and 14-sites lattices $C(j,t)$ shows suppression of the characteristic for CDW "zigzag" structure after quenching. At the same time $P_1(j,t)$ illustrates strong enhancement of the superconducting correlations in nonequilibrium. This on-site correlation function in nonequilibrium shows a quite good agreement with the results for equilibrium superconducting state (dashed lines in Fig.~\ref{fig:scaling}). All in all, for both 10 and 14-sites lattices we found similar behavior after quenching showing the enhancement of the superconducting correlations together with the suppression of the CDW structure.

\addtocounter{figure}{-2}
\begin{figure}[!t]
\centering
%
\includegraphics[width=0.98\columnwidth,clip]{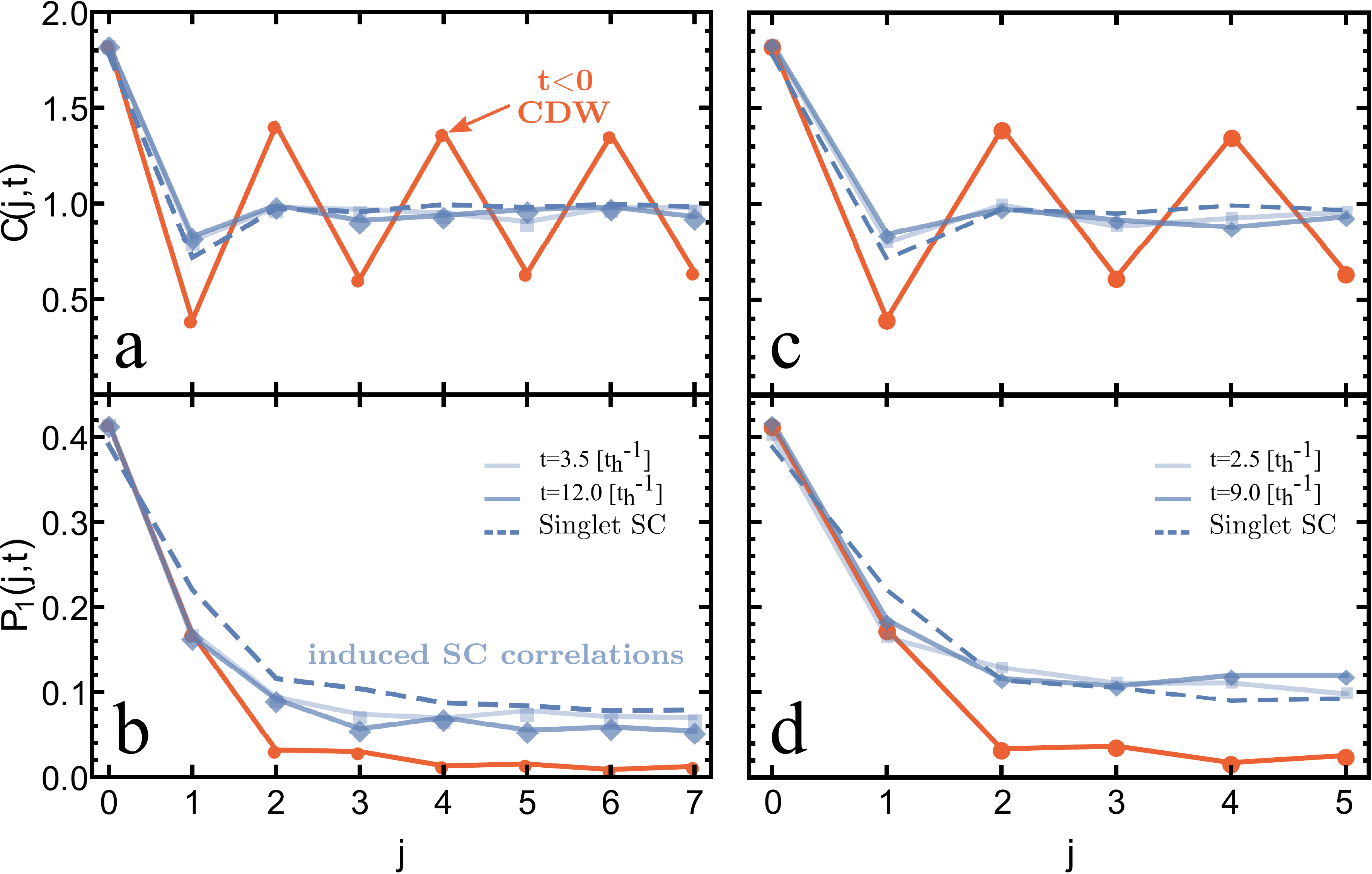}
\caption{(color online) Scaling dependence of (a), (c) the density-density and (b), (d) the on-site correlation 
functions obtained for the interaction quench 
($U=-4$, $V=0.25$, $\Delta V=-0.5$). 
Left panels correspond to the 14-site lattice, and right panels show results for the 10-site lattice. 
The color coding illustrates results at different times: 
red solid line corresponds to $t<0$ , light blue solid line represents $t=3.5$ (14-site) and $t=2.5$ (10-site), and blue solid line is for $t=12$ (14-site) and $t=9$ (10-site). 
Results for equilibrium SC phase are indicated by blue dashed lines. 
}
\label{fig:scaling}
\end{figure}

\addtocounter{figure}{1}
\begin{figure}[!b]
\centering
%
\includegraphics[width=0.85\columnwidth,clip]{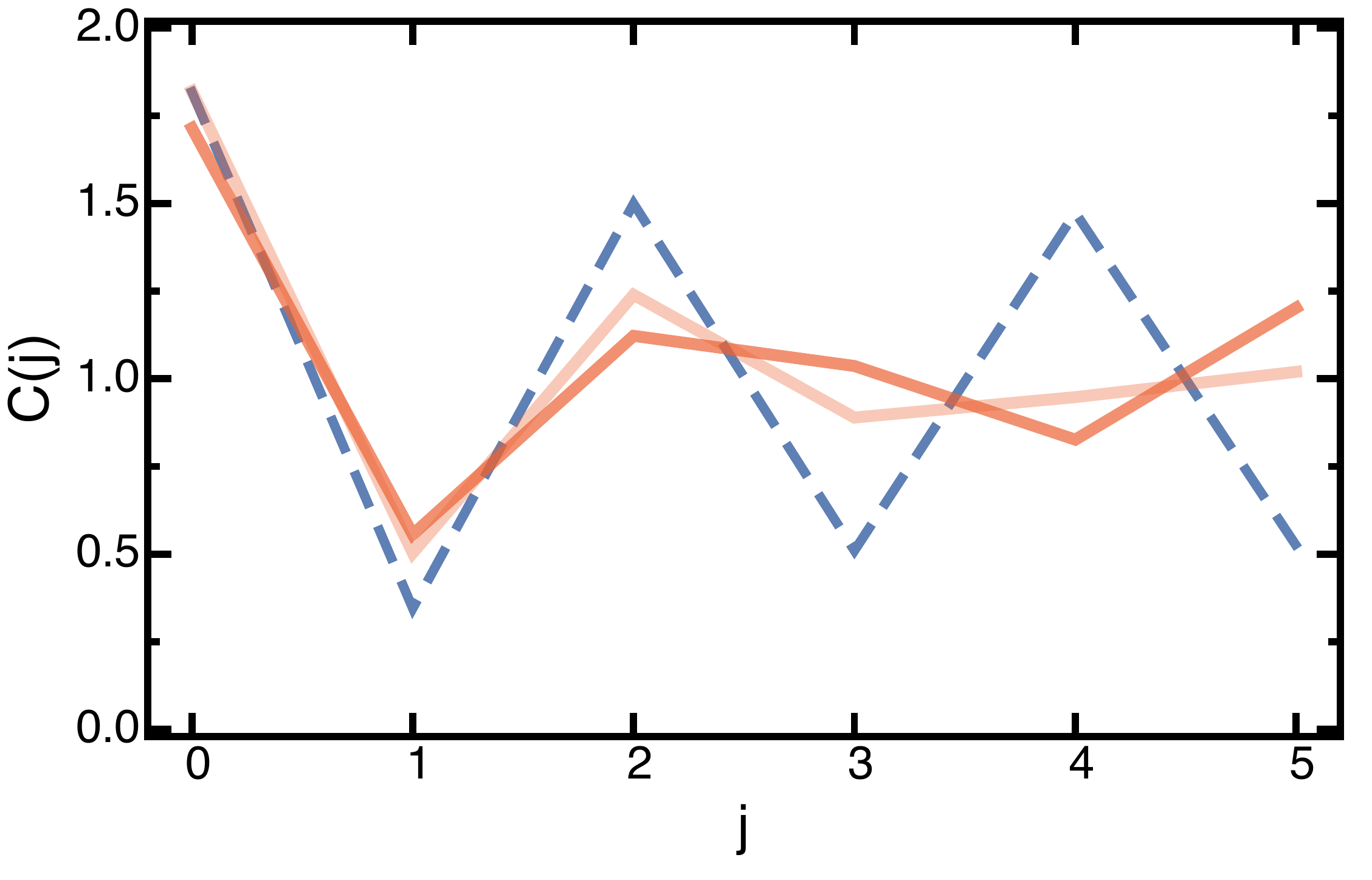}
\caption{(color online) Density-density correlation function versus lattice distance $j$ calculated for $U=-3$ and $V=0.5$.  The calculations are done exemplary for the first 3 eigenstates: ground state with $\Delta E$=0 (blue dashed line), $\Delta E_1=1.6295$ (light red solid line), and $\Delta E_2=2.5479$ (red solid line).
}
\label{fig:C1ExcSt}
\end{figure}

\section*{B. Phase quench.}

\textit{Characterization of eigenstates.} For characterization of the initial phase with $U=-3$ and $V=0.5$ in equilibrium, we plot in Figs.~\ref{fig:P1ExcSt} and~\ref{fig:C1ExcSt} on-site $P_1(j)$  and density-density $C(j)$ correlation functions for several low-energy eigenstates, respectively. Let us focus on the behavior of $C(j)$ (see Fig.~\ref{fig:C1ExcSt}). Whereas in the ground state (blue dashed line) the density-density correlation function shows characteristic "zigzag" structure  indicating alternating order of the electron density with mostly double occupied and empty sites, in the next two low-energy eigenstates (light red and red solid lines) it shows for $j>2$ an almost constant behavior. On the other hand, the on-site correlation function show no significant changes in the low-energy eigenstates (see Fig.~\ref{fig:P1ExcSt}). 
\begin{figure}[!t]
\centering
%
\includegraphics[width=0.98\columnwidth,clip]{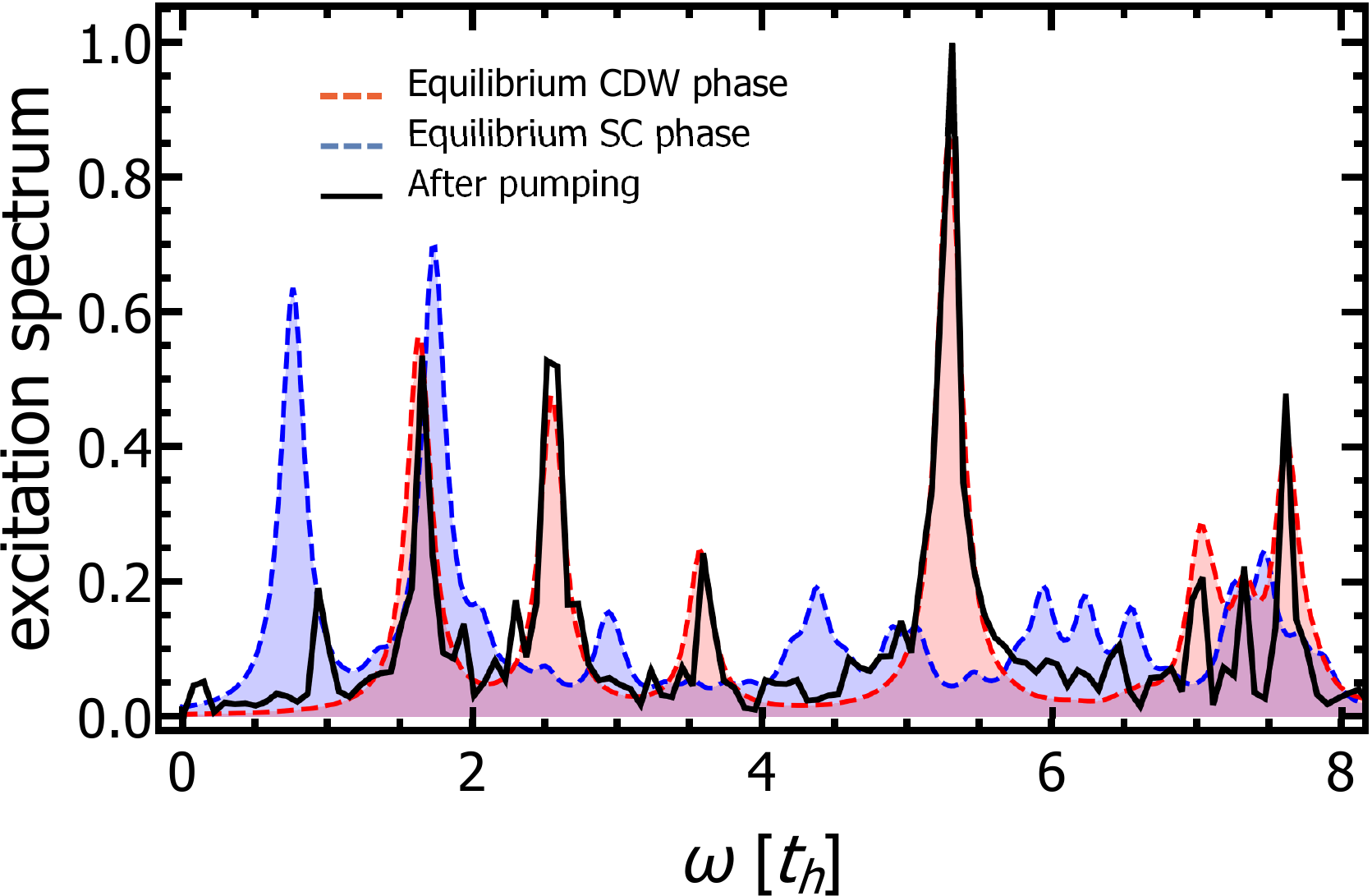}
\caption{(color online) Excitation spectrum of the system in nonequilibrium 
(black solid line) compared with the spectra in equilibrium 
singlet SC phase (blue region) and equilibrium CDW phase (red region). 
}
\label{fig:SM_Fig1}
\end{figure}
\begin{figure}[!b]
\centering
%
\includegraphics[width=0.85\columnwidth,clip]{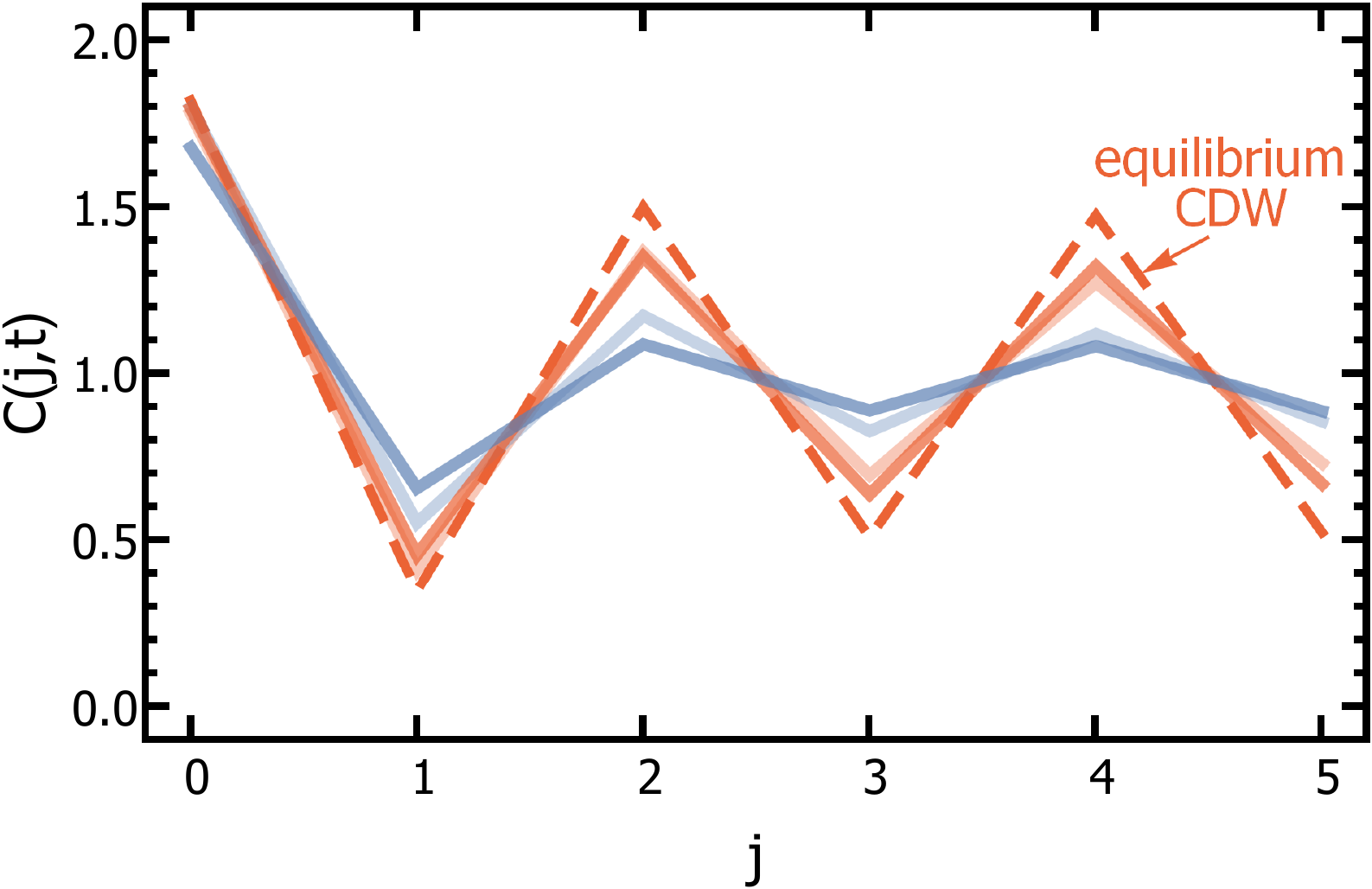}
\caption{(color online) Time--dependent density--density correlation function $C(j, t)$ as 
a function of the lattice site $j$ obtained after a pump pulse excitation of the Gaussian 
shape. The parameters of the pulse are: $A_0=5$, $\omega=2.38$, $\tau=0.05$. The 
function $C(j,t)$ is exemplary shown at times: $t=1$ (light red solid line), 
$t=3$ (light blue solid line), $t=5$ (red solid line), $t=7$ (blue solid line). 
For comparison, the correlation function in equilibrium CDW phase is 
shown by the red dashed line. 
}
\label{fig:SM_Fig2}
\end{figure}

It should be noted that we adjust the central frequency of the pulse used for the phase quench (see main text) to the eigenstate, which is indicated in Figs.~\ref{fig:C1ExcSt} and~\ref{fig:P1ExcSt} by the red solid line.

\textit{Dynamical coexistence. Excitation spectrum and correlation function.}
To calculate an excitation spectrum we use the spectral representation of the stress tensor operator $\hat{\tau}$:
\begin{equation*}
	-\mathrm{Im}\chi_{\tau\tau}(\omega)=\frac{1}{L}\sum_n |\<n|\hat{\tau}|\mathrm{GS}\>|^2 \delta(\omega-(\epsilon_n-\epsilon_0)/\hbar)
\end{equation*}
where $\hat{\tau}=\sum_{j,\sigma}\left(t_h\hat{c}_{j+1,\sigma}^\+\hat{c}_{j,\sigma}+\mathrm{H.c.}\right)$ and $\left.|n\>$ is the $n$-th eigenstate with the energy $\epsilon_n$. For broadening of the spectral lines we use an artificial small number $\eta=1/L$.



In Fig.~\ref{fig:SM_Fig1} we plot the excitation spectrum for initial CDW phase (red line) and equilibrium SC phase (blue line) together with the excitation spectrum after pumping (black line). Latter  
shows several distinct peaks. The most intensive peaks in the 
nonequilibrium spectrum can be identified with the initial equilibrium CDW phase. 
In fact, low energy peaks at $\omega\approx$1.5, 2.5, 3.5 and 5.5 match perfectly
with excitation spectrum for the CDW phase. In addition, a peak at 
$\omega\approx 1.5$ might also represent an excited state of the singlet superconducting 
phase, since one of the most intensive peaks in the optical spectrum of the superconducting 
state appears at the same frequency. Additionally, a peak at $\omega\approx1$ in the 
nonequilibrium spectrum can be assigned to the second intensive peak in the spectrum of the 
superconducting phase. Finally, some less intensive peaks 
at $\omega\approx 5$ and around $\omega\approx 6$ can be assigned to the 
superconducting phase.  

\begin{figure}[!t]
\centering
%
\includegraphics[width=0.8\columnwidth,clip]{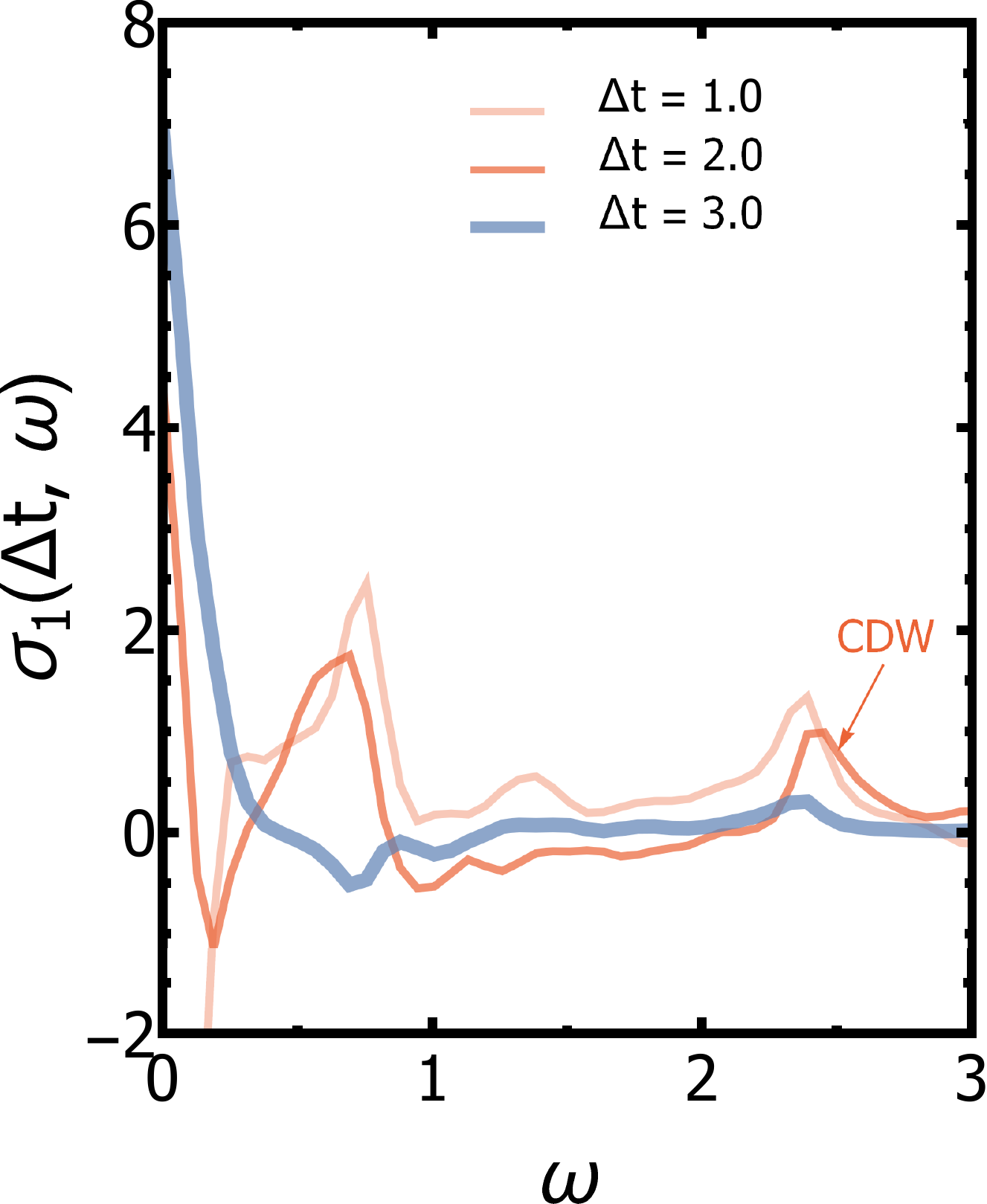}
\caption{(color online) (a) Time--dependent real conductivity $\sigma_1(\Delta t, \omega)$. 
Theoretically obtained results at different time delays are presented by the solid lines: 
$\Delta t$=1 (light red), $\Delta t$=2 (red), and $\Delta t$=3 (blue). 
}
\label{fig:SM_Fig3}
\end{figure}
\begin{figure}[!b]
\centering
\hfill %
\includegraphics[width=0.98\columnwidth,clip]{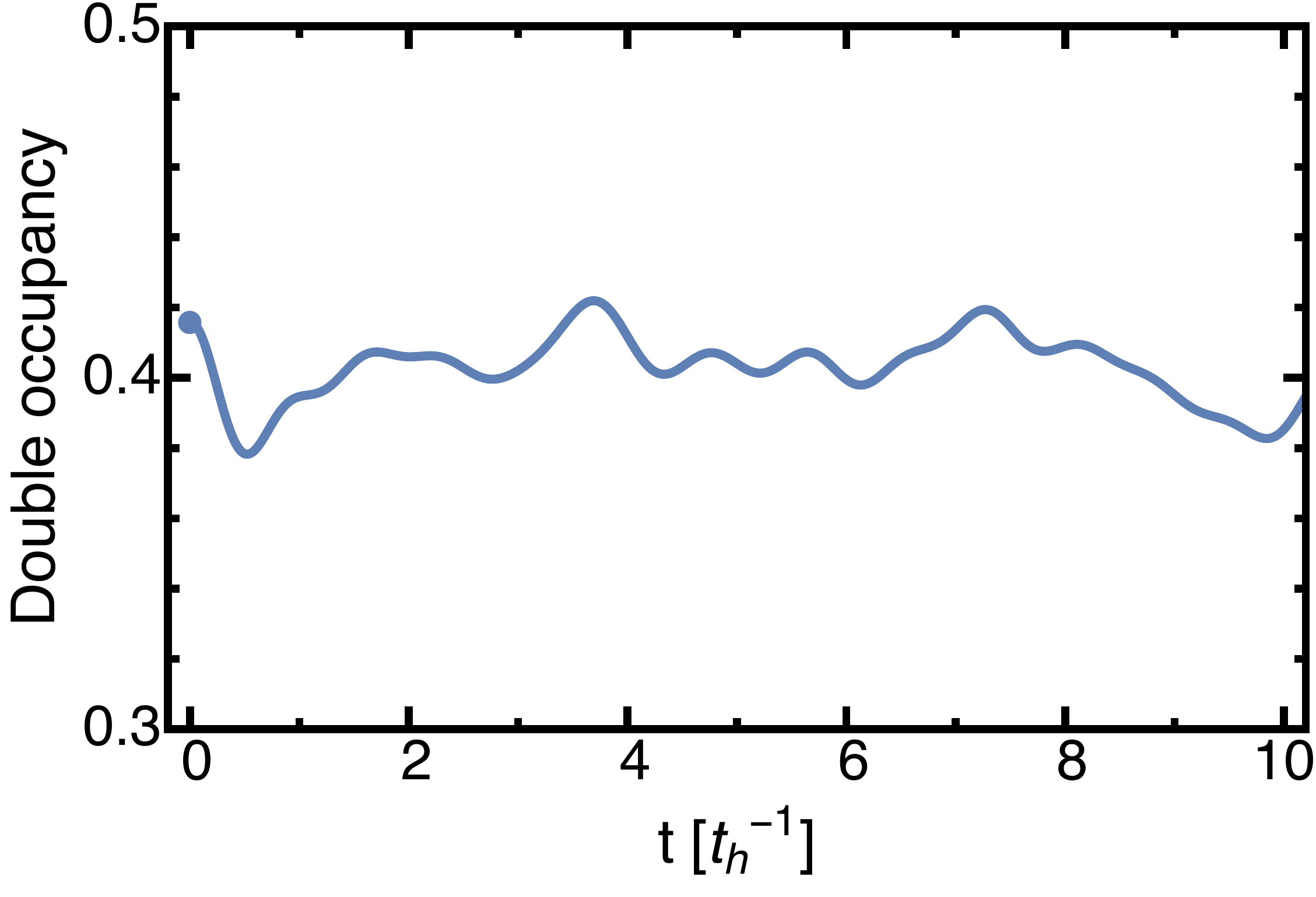}
\caption{(color online) Temporal evolution of the double occupancy function after the interaction quench. Blue dot indicates its value in equilibrium. 
}
\label{fig:DO_quench}
\end{figure}

To explore temporal evolution of the charged order, we plot in Fig.~\ref{fig:SM_Fig2} 
the time--dependent density--density correlation function $C(j,t)$. Before pumping 
the electron system is prepared in the equilibrium ground state of the CDW phase and 
$C(j,t)$ shows a characteristic "zigzag" structure (red dashed line). Clearly, 
the excitation of the system with the pulse leads to an effective partial suppression of the charge density wave correlations with the subsequent oscillations. It should be noted, that the system in nonequlibrium does not indicate any strong similarities with the first low-energy eigenstates of CDW (c.f. Fig.~\ref{fig:C1ExcSt}).

\textit{Optical conductivity.}
Finally, in Fig.~\ref{fig:SM_Fig3} we show the time--dependent real part of 
the optical conductivity $\sigma_1(\Delta t, \omega)$. 
In agreement with the results for the imaginary conductivity $\sigma_2(\Delta t, \omega)$ [see Fig. 5 in the main text] we observe first a partial reduction of the spectral weight of 
the low energy peak corresponding to the equilibrium CDW state at both $\Delta t=1$ 
(light red solid line) and $\Delta t=2$ (red solid line) with the appearance of an in--gap state at $\omega\approx 0.7$. At $\Delta t=3$ the response from CDW is disappeared and the 
spectral weight is shifted to a low--energy peak at $\omega\approx 0$ 
representing a transient $\d$--peak.

\begin{figure}[!b]
\centering
\hfill %
\includegraphics[width=0.98\columnwidth,clip]{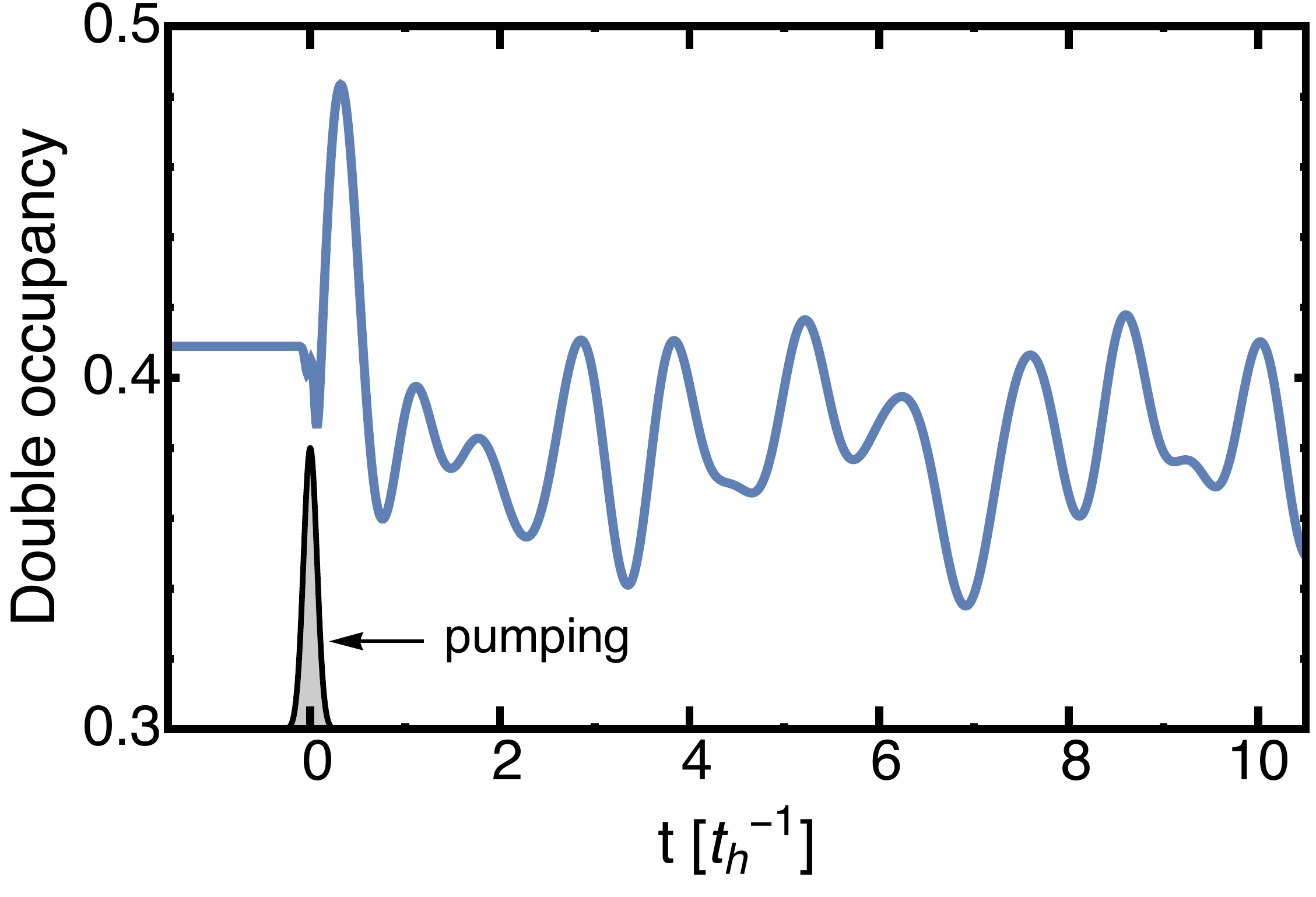}
\singlespacing
\caption{(color online) Temporal evolution of the double occupancy function after the phase quench. 
}
\label{fig:DO_pulse}
\end{figure}
\section*{C. Double occupancy function}

In order to get an additional insight into the charge dynamics after the interaction and the phase quench we plot in Figs.~\ref{fig:DO_quench} and~\ref{fig:DO_pulse} time-dependent double occupation function. After the  interaction quench double occupation function
shows oscillations with a quite small magnitude (see Fig.~\ref{fig:DO_quench}). Physically this behavior can be interpreted as a redistribution of the electron pairs on the lattice initially prepared in the CDW phase, which was initiated by the interaction quench. 

In case of the pulse quench (see Fig.~\ref{fig:DO_pulse}) we find in the behavior of the double occupancy function an effective decrease with strong oscillations around a new reduced value. Physically, it means that the excitation of the electron system leads to a dynamical breaking and creation of the electron pairs on a lattice. 
Moreover, at some moments in time (e.g. at $t \approx 3$ etc.) one can observe a recovery of the double occupation function to its initial value. Since the charge density wave correlations are partially suppressed after pumping (see Fig.~\ref{fig:SM_Fig2}), this behavior indicates a transient redistribution of the electron pairs on the lattice.

\end{document}